\documentclass[conference]{IEEEtran}
\IEEEoverridecommandlockouts
\usepackage{amsmath,amssymb,amsfonts}
\usepackage{algorithmic}
\usepackage{graphicx}
\usepackage{textcomp}
\usepackage{xcolor}
\usepackage{booktabs}
\usepackage{float}
\usepackage{caption}
\usepackage{placeins}
\usepackage{svg}
\usepackage{multirow}
\usepackage[utf8]{inputenc}
\usepackage{hyperref}

\def\BibTeX{{\rm B\kern-.05em{\sc i\kern-.025em b}\kern-.08em
    T\kern-.1667em\lower.7ex\hbox{E}\kern-.125emX}}
    
\begin{document}

\bstctlcite{IEEEexample:BSTcontrol}

\title{Teager-Kaiser Energy Methods for EEG Feature Extraction in Biomedical Applications%
\thanks{
This project is funded by the European Union under Horizon Europe (grant No. 101136568 - HERON).\\
\includegraphics[height=2.5em]{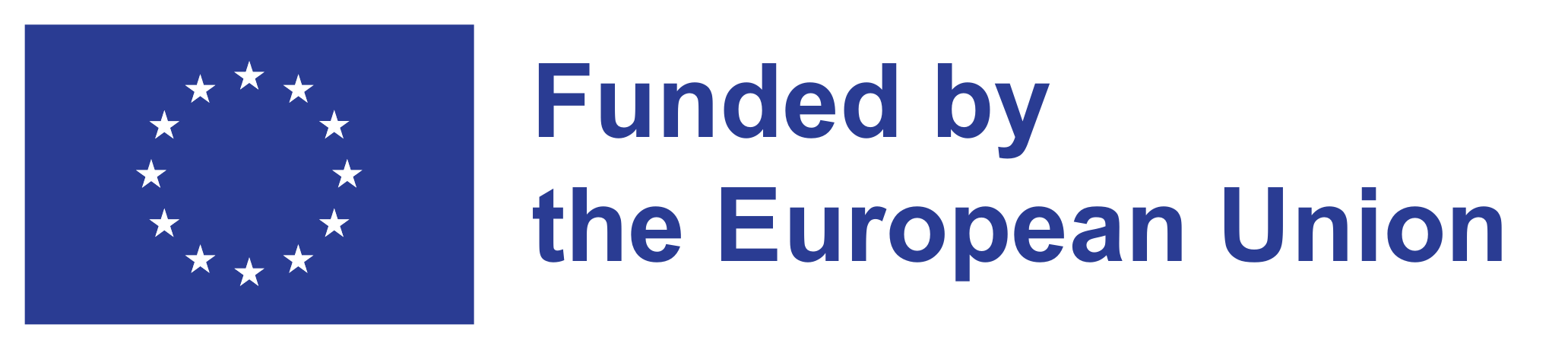}
}}


\author{\IEEEauthorblockN{Ioanna Chourdaki\IEEEauthorrefmark{1},
Kleanthis Avramidis\IEEEauthorrefmark{4},
Christos Garoufis\IEEEauthorrefmark{1}\IEEEauthorrefmark{2}\IEEEauthorrefmark{3}, 
Athanasia Zlatintsi\IEEEauthorrefmark{1}\IEEEauthorrefmark{2}\IEEEauthorrefmark{3}, and
Petros Maragos\IEEEauthorrefmark{1}\IEEEauthorrefmark{2}\IEEEauthorrefmark{3}}
\IEEEauthorblockA{\IEEEauthorrefmark{1}School of ECE, National Technical University of Athens, Athens, Greece}
\IEEEauthorblockA{\IEEEauthorrefmark{2}Institute of Robotics, Athena Research Center, Maroussi, Greece}
\IEEEauthorblockA{\IEEEauthorrefmark{3}HERON - Hellenic Robotics Center of Excellence, Athens, Greece}
\IEEEauthorblockA{\IEEEauthorrefmark{4}Signal Analysis and Interpretation Lab, University of Southern California, USA}
}
\maketitle

\begin{abstract}
Electroencephalography (EEG) signals are inherently non-linear, non-stationary, and vulnerable to noise sources, making the extraction of discriminative features a long-standing challenge. In this work, we investigate the non-linear Teager-Kaiser Energy Operator (TKEO) for modeling the underlying energy dynamics of EEG in three representative tasks: motor imagery, emotion recognition, and epilepsy detection. To accommodate the narrowband nature of the operator, we employ Gabor filterbanks to isolate canonical frequency bands, followed by the Energy Separation Algorithm to decompose the TKEO output into an amplitude envelope and instantaneous frequency components. We then derive a set of energy descriptors based on this demodulation and compare their classification performance against established EEG features. The proposed TKEO-based pipeline offers an intuitive, physiologically grounded framework for capturing EEG signal dynamics, while remaining simple, training-free, and data-efficient. Our findings suggest that combining TKEO features with conventional ones improves Balanced Accuracy by approximately 15\% in epilepsy detection, yields modest gains in motor imagery, and achieves on par performance in emotion recognition, reflecting the pipeline’s ability to capture transient neural dynamics.

\end{abstract}

\begin{IEEEkeywords}
Teager-Kaiser Energy Operator, Amplitude-Frequency Modulation, EEG, Biomedical Applications
\end{IEEEkeywords}

\section{Introduction}
\label{sec:intro}
Human brain activity underlies critical cognitive and affective processes, yet many aspects of its neural mechanisms are not adequately understood. Among the available recording choices, Electroencephalography (EEG) is particularly popular, as it offers cost-effective, minimally invasive means of capturing oscillatory neural dynamics at high temporal resolution. At the same time, EEG signals are high-dimensional and notoriously noisy, making the extraction of informative features a persistent challenge. Despite the success of deep learning approaches in other domains, analysis of biosignals like EEG is ultimately bounded by limitations in data availability and subject heterogeneity, which has led many studies to continue relying on handcrafted, domain-specific features rather than purely data-driven representations~\cite{malekzadeh2021}. Conventional spectral and energy-based measures mainly capture overt oscillatory power~\cite{psd1}, often overlooking the transient and non-linear dynamics that govern several neurophysiological behaviors.


Traditional EEG analysis relies on handcrafted features capturing statistical, spectral, and dynamical signal properties. Common descriptors include basic statistics~\cite{EEG-featutres_stats}, with spectral features --particularly Power Spectral Density (PSD)-- remaining the most common approach for quantifying canonical oscillatory rhythms~\cite{psd1}. Additional measures include differential entropy~\cite{invastigating_critical}, asymmetry indices~\cite{EEG-featutres_asymmetry}, and Hilbert Transform (HT)-based representations~\cite{hilbert_transform} for temporal and spatial complexity. Multiscale and non-stationary dynamics are typically modeled using Discrete Wavelet Transform (DWT) and Empirical Mode Decomposition (EMD)~\cite{emd_dwt}, while non-linear descriptors such as fractal dimensions~\cite{avramidis2021} capture more complex EEG behavior. Despite this rich toolbox, most approaches rarely target an explicit analysis of transient and non-linear dynamics in EEG.

To address this limitation, the Teager-Kaiser Energy Operator (TKEO)~\cite{kaiser, maragos1993} provides a complementary approach. As a non-linear differential operator, TKEO yields instantaneous energy by estimating modulations of both amplitude and dominant frequencies in oscillatory signals. This makes it well-suited for EEG analysis, where recorded oscillatory activity can be naturally decomposed into frequency bands with distinct clinical functions~\cite{eeg_basics_book}. 

Building on these properties, TKEO has been employed across signal processing domains, with early applications in speech and music processing for tasks such as formant detection~\cite{tkeo_formants} and music instrument recognition~\cite{zlatintsi2012}. Beyond acoustic signals, TKEO has proven effective in biomedical applications to extract physiologically plausible energy fluctuations. In cardiovascular research, for example, it has enabled automated phonocardiogram analysis for heart valve disorder detection without prior segmentation~\cite{detection-of-heart-disorder}. In neurophysiological research, it has been applied both on electromyography and electroencephalography, supporting applications such as movement onset detection~\cite{upper_limb_movement} and focal activity detection~\cite{focal}.


In this study, we systematically evaluate TKEO-based features across two tasks characterized by well-established transient dynamics --motor imagery and epilepsy detection-- as well as in emotion recognition, which tests the method’s applicability beyond its original motivating context. Through this evaluation, we clarify the strengths and limitations of TKEO features in EEG analysis and identify conditions under which they provide added value. While end-to-end deep learning is increasingly applied to EEG analysis, our methodology emphasizes model comparisons and feature interpretability, elements that are obscured and difficult to quantify in end-to-end training. TKEO-based energy descriptors instead improve the performance of conventional features in motor imagery and epilepsy detection while still performing on par in emotion recognition. 
Our contribution can be summarized as follows:

\begin{itemize}
    \item We motivate and extract a comprehensive set of EEG features derived from instantaneous energy analysis using the TKEO.
    \item We introduce a systematic evaluation of TKEO-derived features against conventional PSD-based energy, squared-energy features, HT-based features, and differential entropy, across three representative EEG benchmarks.
    \item We provide empirical evidence of the conditions under which TKEO features outperform and/or yield complementary insights over alternative energy measures, particularly in tasks with transient and non-linear dynamics.
\end{itemize}

\section{Methodology}
\label{sec:pagestyle}

\subsection{Teager-Kaiser Energy Operator}
\label{ssec:tkeo}
Many physiological signals exhibit oscillatory electrical activity that can be described in terms of amplitude and frequency modulations (AM-FM). To capture such modulations, TKEO was introduced as a non-linear differential operator that estimates the instantaneous energy of a signal \cite{kaiser,maragos1993}. This operator estimates the instantaneous energy of a signal as a function of its time-varying amplitude and frequency, effectively serving as a \textit{tracker} of energy variations. TKEO was originally formulated for both continuous and discrete signals. For a continuous signal $x(t)$ it is defined as
\begin{equation}
    \Psi_c[x(t)] = (\dot{x}(t))^2-x(t)\,\ddot{x}(t),
\end{equation}
while for a discrete-time signal $x[n]$ it takes the form
\begin{equation}
    \Psi_d[x[n]] = x^2[n]-x[n-1]\,x[n+1].
\end{equation}
For the purpose of this study, we constrain our analysis to the discrete-time TKEO formulation, i.e., $\Psi \equiv \Psi_d$.

\subsection{Filterbank Decomposition}
\label{ssec:filterbanks}
TKEO provides reliable instantaneous energy estimates primarily when applied to narrowband signals~\cite{maragos1993}. To apply this premise, the use of filterbanks helps decompose the broadband (raw) signal into localized frequency subbands, which, as a result, produces smoother instantaneous amplitude and frequency modulations \cite{dimitriadis2004}. Among possible choices, Gabor filters are particularly suitable due to their joint time-frequency resolution and compactness~\cite{maragos1993}.

A Gabor filterbank is constructed by generating a sequence of Gabor filters whose center frequencies are uniformly distributed between a lower ($f_{\mathrm{low}}$) and an upper ($f_{\mathrm{high}}$) cutoff frequency. In this way, the broadband signal is decomposed into subbands, each filter capturing localized spectral content around its center frequency. In the EEG setting, we apply this procedure to both the broadband signal (0.5-100 Hz) and each canonical frequency band: Delta (0.5-3 Hz), Theta (4-7 Hz), Alpha (8-12 Hz), Beta (13-30 Hz), and Gamma (30-50 Hz). For each band, $f_{\mathrm{low}}$ and $f_{\mathrm{high}}$ are set to the band boundaries, and $N$ filters are uniformly distributed inside this interval. The frequency step is defined as: $\Delta f = \frac{f_{\mathrm{high}}-f_{\mathrm{low}}}{N}$, where $N$ is the number of filters. The center frequencies are then: $f_\mathrm{c}(k) = f_{\mathrm{low}} + k\Delta f$, where $k \in \mathbb{Z}_{\geq 0}:k < N$. Once the set of $f_c$ has been determined, each EEG signal is filtered with the corresponding Gabor filters, and TKEO is applied to each filter's output, resulting in $N$ energy signals, one per subband.

For each TKEO signal, we compute the temporal mean as a measure of the average subband energy. Finally, for both the broadband and each canonical band, the filter of maximum averaged TKEO value is selected, and the corresponding narrowband signal is retained for further analysis, as it represents the subband in which the TKEO provides the most informative characterization of the signal dynamics.


\subsection{Energy Separation Algorithm}
\label{ssec:desa}
\begin{figure}[t]
\centering
\includegraphics[width=\linewidth]{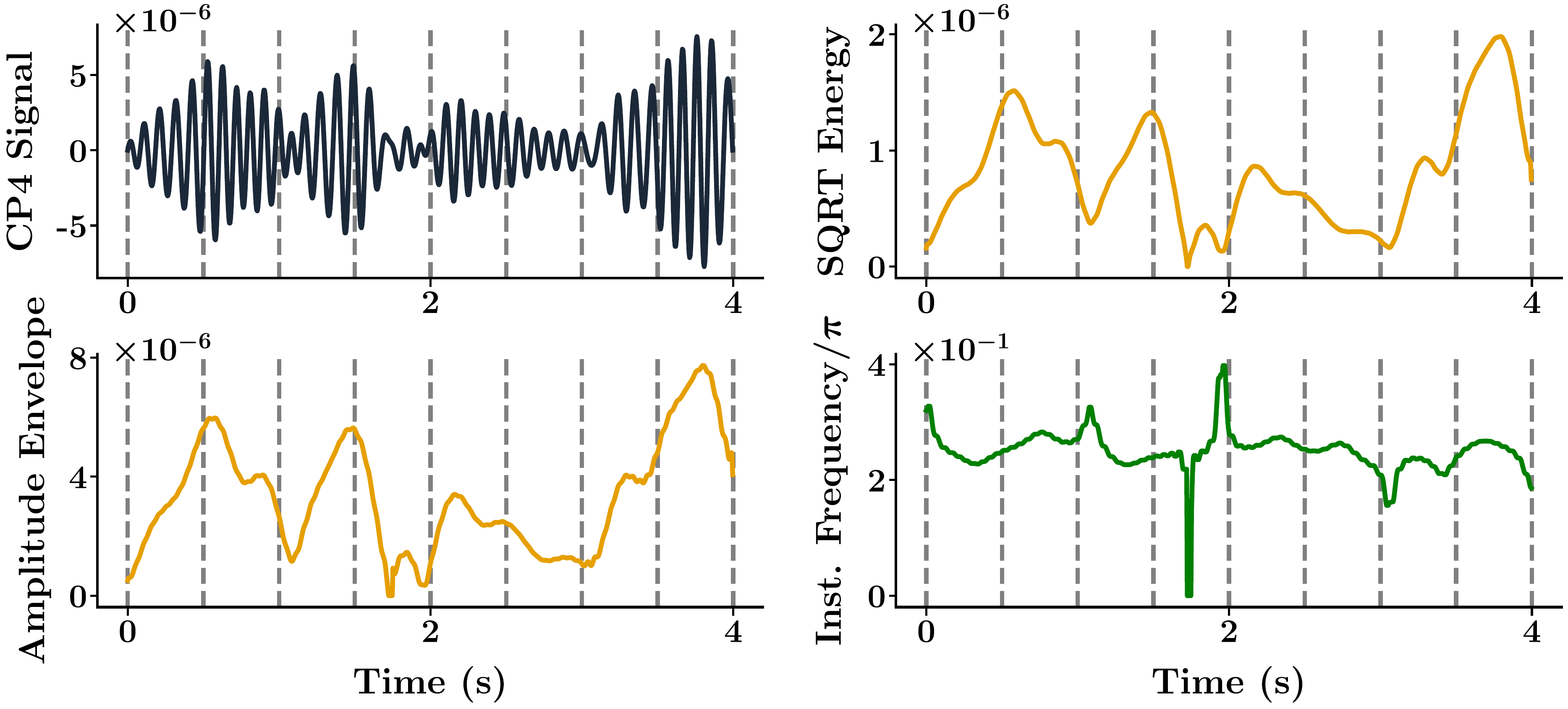}
\caption{(a) CP4 channel signal from the BCI-IV\,2a dataset in the Alpha band. (b) Square root of TKEO. (c) Estimated amplitude envelope using DESA-1. (d) Estimated instantaneous frequency using DESA-1, expressed as a fraction of $\pi$.}
\label{fig:res}
\vspace{-0.4cm}
\end{figure}

To estimate the amplitude envelope and instantaneous frequency of AM-FM signals, Maragos et al.~\cite{maragos1993} introduced the Energy Separation Algorithm (ESA), which employs non-linear combinations of TKEO outputs to decouple amplitude and frequency components. For a discrete-time signal $x[n]$, this framework was extended to the Discrete Energy Separation Algorithm (DESA)~\cite{maragos1993}. Here we adopt the \mbox{DESA-1} variant, for which the amplitude envelope $|a[n]|$ and the instantaneous frequency $\Omega[n]$ of $x[n]$ are defined as
\begin{equation}
    |a[n]| \approx \sqrt{\frac{\Psi[x[n]]}{1 - \left[1 - \frac{\Psi\big[y[n]\big]+\Psi\big[y[n+1]\big]}{4\,\Psi[x[n]]}\right]^2}},
\end{equation}
\begin{equation}
    \Omega[n] \approx \arccos\!\left(1 - \frac{\Psi\big[y[n]\big]+\Psi\big[y[n+1]\big]}{4\,\Psi[x[n]]}\right),
\end{equation}
where $y[n] = x[n] - x[n-1]$. An example of estimating the amplitude envelope and the instantaneous frequency from an EEG signal is depicted in Fig.~\ref{fig:res}. These quantities are subsequently used for feature extraction, as described in Sec.~\ref{ssec:tkeo_features}. 


\subsection{TKEO Feature Extraction}
\label{ssec:tkeo_features}
We extract TKEO-based features from each EEG channel across either the predefined frequency bands, or the broadband signal. We first compute the \textit{mean Teager-Kaiser energy (\mbox{m-TKEO})}. For a band-limited signal $s$, the \textit{relative} contribution of its Teager-Kaiser energy with respect to all frequency bands under consideration is:
\begin{equation}
    \text{RE}_{\,\text{band}}(s) = \frac{\Psi_{\text{band}}(s)}{\sum_{k} \Psi_{k}(s)},
\end{equation}
where band $k\in\{\text{Delta},\,\text{Theta},\,\text{Alpha},\,\text{Beta},\,\text{Gamma}\}$. Then, we calculate the \textit{mean Relative Energy (m-RE)}. Finally, applying DESA-1, we obtain the instantaneous amplitude $|a[n]|$ and frequency $\Omega[n]$, from which we compute the \textit{mean Instantaneous Amplitude Modulation (m-IAM)}~\cite{zlatintsi2012} and the \textit{variance of Instantaneous Frequency Modulation (\mbox{v-IFM})}. From this point onward, by TKEO features we refer to the following 4D feature set: m-TKEO, m-RE, m-IAM, and \mbox{v-IFM} features.

\section{Experimental Setup}
\label{sec:setup}
\vspace{-0.05cm}
\subsection{Datasets}
\textbf{BCI Competition IV Dataset 2a:} The BCI-IV\,2a dataset~\cite{bci2a} consists of EEG recordings from 9 subjects performing 4 motor imagery tasks, i.e., left-hand, right-hand, feet, and tongue movement imagery. Each subject completed two sessions on different days, with each session comprising 6 runs of 48 trials (12 per class), resulting in 288 trials per session. Each trial lasted 6 seconds in total, with an onset cue presented after the first 2 seconds. Subjects performed the instructed motor imagery for the subsequent 4 seconds, which was  used as the analysis window. EEG was acquired from 22 channels following the 10-20 configuration~\cite{10-20system}, along with 3 electrooculogram (EOG) channels, all sampled at 250 Hz.



\textbf{SEED:} The SEED dataset~\cite{invastigating_critical,seed2} is a widely used benchmark for EEG-based emotion recognition. It comprises EEG recordings from 15 participants while they viewed 15 Chinese film clips, each lasting approximately 4 minutes. Each video was selected to induce one of three affective states (positive, neutral or negative) resulting in a single categorical label per clip. The EEG signals were recorded using a 62-channel setup following the 10-20 configuration, with an original rate of 1000\,\,Hz, subsequently downsampled to 200\,\,Hz.

\textbf{TUH Epilepsy Corpus:} The TUH Epilepsy Corpus (TUEP)~\cite{tuh} is a curated collection of clinical EEG data from Temple University Hospital, including recordings from matched 100 epileptic and 100 healthy subjects. In total, TUEP contains 698 sessions, yielding 1,785 EEG recordings from epileptic and 513 from non-epileptic patients. Individual recording durations vary from a few seconds to approximately one hour. All EEG signals were acquired using the \mbox{10-20} configuration with 19 channels sampled at 250 Hz with supplementary electrocardiogram (EKG) channels available.

\subsection{Preprocessing}
All recordings were segmented into fixed-length windows using dataset-specific lengths and overlap ratios chosen according to the characteristics and experimental protocols of each dataset. For BCI-IV\,2a, the 4s motor imagery interval per trial was used, SEED was segmented into 20s windows with 50\% overlap, and TUEP into 10s non-overlapping windows. SEED was provided in a preprocessed format, which included manual removal of EMG and EOG artifacts, notch filtering at 50\,Hz and lowpass filtering at 75\,Hz~\cite{invastigating_critical}. For the other two datasets, we applied a similar pipeline that involved Independent Component Analysis (ICA) to remove EKG artifacts in TUEP and EOG artifacts in BCI-IV\,2a. Then, notch filtering was applied at 60~Hz for TUEP and 50~Hz for BCI-IV\,2a to eliminate powerline interference. Finally, all signals were highpass-filtered at 0.5~Hz.

Before TKEO-based feature extraction, each EEG segment was decomposed into five narrowband components using bandpass filtering corresponding to the canonical EEG bands. Within both the frequency range of each canonical band and the raw signal we applied a filterbank of 12 filters, as per~\cite{zlatintsi2012}.



\vspace{-0.1cm}
\subsection{Baseline Feature Extraction}
To enable a comparative evaluation of the proposed approach and assess its potential complementarity with established EEG descriptors, we re-implement a standard feature extraction pipeline for EEG signals, wherein we first isolate the desired frequency bands with a 10th-order Butterworth filter, and then extract a set of baseline features from the isolated bands (or the raw signal). These features include the \textit{Power Spectral Density (PSD)}, the \textit{Signal Energy (SE)}, the \textit{Hilbert Transform (HT)}, and the \textit{Differential Entropy (DE)}.
PSD is computed using the Welch's method, yielding $f_s/2$ features per signal, where $f_s$ denotes the sampling frequency. The SE, defined as the squared magnitude of the raw signal at any point in time and commonly used in signal processing as a direct measure of instantaneous energy, is employed to derive two energy-based features: the \textit{mean Signal Energy} and the \textit{mean Relative Signal Energy}, analogous to the proposed TKEO statistical descriptors. The HT is applied to obtain the analytic signal, from which instantaneous amplitude and instantaneous frequency are derived as the magnitude of the analytic signal and the time derivative of the unwrapped phase scaled by $f_s$, respectively. Subsequently, the \textit{mean Instantaneous Amplitude} and the \textit{variance of the Instantaneous Frequency} are extracted.  Finally, for a band-limited signal $s$, the DE is defined as $h(s)=\frac{1}{2}\log(2 \pi e \sigma_s^2)$, where $\sigma_s^2$ the signal variance~\cite{invastigating_critical}. For clarity, we define the \textit{TKEO+X} features as the TKEO features concatenated with $\text{X}\in\{\text{SE}, \text{PSD}, \text{HT}, \text{DE}\}$ features.


\vspace{-0.1cm}
\subsection{Evaluation Protocol} 
We evaluate the utility of TKEO-based features on three classification tasks: motor imagery (BCI-IV\,2a dataset), emotion recognition (SEED dataset) and epilepsy detection (TUEP dataset). A \textit{Subject Independent} setting is presented, where models are trained on a subset of subjects and tested on unseen individuals. This setting reduces the risk of confounding due to subject- or recording-specific biases and better assesses cross-subject generalization, better reflecting real-world applications.

Features from all EEG channels were concatenated and the resulting feature vectors were standardized with a Standard Scaler and classified using a Random Forest classifier with 100 estimators, chosen for computational efficiency. We deliberately employed Random Forest, as it is a standard classifier that allows us to assess the discriminative power of the extracted features. This choice ensures that classification performance reflects the quality of the proposed TKEO-based features, rather than the capacity of complex machine learning models. Evaluation is performed with stratified 5-fold cross-validation. SEED and BCI-IV\,2a are balanced, whereas TUEP is imbalanced, with 77.7\% epileptic recordings; to mitigate this, we report metrics robust to class imbalance, namely Balanced Accuracy\,and ROC-AUC,\,following prior\,literature\,\cite{metrics_literature}.

\section{Results \& Discussion}
\label{sec:experiments}

\textbf{Main Results: }The classification results for all features are summarized in Tables~\ref{tab:results_mi}-\ref{tab:results_tuep}. We report results using both features extracted directly from the raw signal (\textit{raw-band}) and from the fusion of features computed across all canonical frequency bands (\textit{fused-band}). TKEO features show competitive performance to the baselines, with modest gains in motor imagery and a substantial complementary effect in epilepsy detection when combined with them; in emotion recognition, TKEO-based features perform on par with the baselines.


As shown in Table~\ref{tab:results_mi}, TKEO features consistently outperform all baselines in motor imagery. Moreover, integrating TKEO features with the baselines improves their performance in all cases, highlighting their complementary nature, with the fused-band achieving the best overall performance (41.0\% Balanced Accuracy and 67.1\% ROC-AUC) -- substantially above chance level. In the SEED dataset (Table~\ref{tab:results_seed}), TKEO features alone achieve performance comparable to the baselines. Integrating TKEO with other features mostly enhances classification results, with the TKEO\,{+}\,SE set achieving the highest performance (69.0\% Balanced Accuracy, 85.5\% ROC-AUC). In the TUEP dataset (Table~\ref{tab:results_tuep}), combining the TKEO features with PSD (TKEO\,{+}\,PSD) achieves the highest performance, with the fused-band configuration achieving the highest Balanced Accuracy (75.9\%)--approximately 15\% higher on average than other feature sets--and the highest ROC-AUC (80.7\%). TKEO features alone attain performance comparable to the baselines, further highlighting their complementary contribution. In the epilepsy detection task, this complementarity is intuitive: TKEO features capture transient modulations, whereas PSD characterizes the distribution of spectral power. Since epileptic EEG contains both transient non-stationary events and sustained rhythmic activity~\cite{eeg_basics_book}, these features may provide complementary information.

Overall, TKEO features show stronger relative performance in motor imagery and epilepsy detection compared to emotion recognition, underscoring their suitability for tasks where transient temporal changes are important and their limitations when the discriminative information is not as temporally localized. Additionally, the fused-band outperforms the \mbox{raw-band} in most cases within TKEO features, reflecting the advantage of aggregating information across frequency bands while preserving the narrowband characteristics of TKEO. The proposed framework remains training-free, as the additional processing stages do not introduce trainable parameters, while offering interpretable descriptors of transient EEG dynamics that appear complementary to conventional features.


\begin{table}[!h]
\fontsize{7.8pt}{8.8pt}\selectfont
\setlength{\tabcolsep}{3pt}
\renewcommand{\arraystretch}{0.95}
\captionsetup{skip=-2pt}
\caption{5-fold Mean Classification Performance and Standard Deviation on BCI-IV\,2a (Motor Imagery).}
\begin{center}
\begin{tabular}{|c|c|c|c|c|}
\hline
\multirow{2}{*}{\textbf{Features}} 
& \multicolumn{2}{|c|}{\textbf{Balanced Accuracy (\%)}} 
& \multicolumn{2}{|c|}{\textbf{ROC-AUC ($\mathbf{\times 10^2}$)}} \\
\cline{2-5}
 & \textbf{Raw} & \textbf{Fused} & \textbf{Raw} & \textbf{Fused} \\
\hline
SE & 33.9$_{\,\pm3.7}$ & 37.6$_{\,\pm5.2}$ & 61.4$_{\,\pm4.2}$ & 64.3$_{\,\pm4.9}$ \\
PSD & 37.8$_{\,\pm6.7}$ & 35.7$_{\,\pm4.6}$ & 64.5$_{\,\pm6.4}$ & 62.8$_{\,\pm5.4}$ \\
HT & 36.2$_{\,\pm5.1}$ & 36.9$_{\,\pm5.6}$ & 62.4$_{\,\pm4.5}$ & 64.6$_{\,\pm4.9}$ \\
DE & 35.0$_{\,\pm4.7}$ & 39.3$_{\,\pm5.1}$ & 61.9$_{\,\pm4.5}$ & 65.1$_{\,\pm5.6}$ \\
\hline
TKEO & \textbf{40.4$_{\,\pm7.0}$} & 39.9$_{\,\pm7.3}$ & 65.9$_{\,\pm7.1}$ & 66.5$_{\,\pm6.3}$  \\
TKEO\,{+}\,SE & 39.1$_{\,\pm6.2}$ & 38.8$_{\,\pm6.6}$ & 65.4$_{\,\pm5.6}$ & 66.5$_{\,\pm6.4}$ \\
TKEO\,{+}\,PSD & 40.3$_{\,\pm5.2}$ & 38.3$_{\,\pm6.8}$ & \textbf{66.2$_{\,\pm5.9}$} & 64.3$_{\,\pm5.5}$ \\
TKEO\,{+}\,HT & 39.8$_{\,\pm5.8}$ & \textbf{41.0$_{\,\pm7.2}$} & 66.1$_{\,\pm5.8}$ & \textbf{67.1$_{\,\pm6.2}$} \\
TKEO\,{+}\,DE & 40.1$_{\,\pm7.0}$ & 40.5$_{\,\pm8.1}$ & 66.0$_{\,\pm6.1}$ & 66.8$_{\,\pm6.4}$ \\
\hline
\end{tabular}
\label{tab:results_mi}
\end{center}
\vspace{-0.7cm}
\end{table}


\begin{table}[!h]
\fontsize{7.8pt}{8.8pt}\selectfont
\setlength{\tabcolsep}{3pt}
\renewcommand{\arraystretch}{0.95}
\captionsetup{skip=-2pt}
\caption{5-fold Mean Classification Performance and Standard Deviation on SEED (Emotion Recognition).}
\begin{center}
\begin{tabular}{|c|c|c|c|c|}
\hline
\multirow{2}{*}{\textbf{Features}} 
& \multicolumn{2}{|c|}{\textbf{Balanced Accuracy (\%)}} 
& \multicolumn{2}{|c|}{\textbf{ROC-AUC ($\mathbf{\times 10^2}$)}} \\
\cline{2-5}
 & \textbf{Raw} & \textbf{Fused} & \textbf{Raw} & \textbf{Fused} \\
\hline
SE & 55.4$_{\,\pm4.5}$ & 68.5$_{\,\pm3.9}$ & 73.4$_{\,\pm4.4}$ & 85.0$_{\,\pm2.9}$ \\
PSD & \textbf{63.9$_{\,\pm1.4}$} & 62.5$_{\,\pm3.9}$ & \textbf{81.9$_{\,\pm1.6}$} & 80.7$_{\,\pm3.8}$ \\
HT & 62.3$_{\,\pm1.9}$ & 68.2$_{\,\pm4.4}$ & 80.1$_{\,\pm2.1}$ & 84.7$_{\,\pm3.5}$ \\
DE & 52.5$_{\,\pm4.3}$ & 67.2$_{\,\pm3.5}$ & 70.9$_{\,\pm4.0}$ & 84.4$_{\,\pm2.9}$ \\
\hline 
TKEO & 61.7$_{\,\pm3.3}$ & 67.3$_{\,\pm4.0}$ & 79.7$_{\,\pm2.7}$ & 84.1$_{\,\pm2.9}$  \\
TKEO\,{+}\,SE & 62.5$_{\,\pm3.4}$ & \textbf{69.0$_{\,\pm3.9}$} & 80.7$_{\,\pm2.6}$ & \textbf{85.5$_{\,\pm2.7}$} \\
TKEO\,{+}\,PSD & 63.3$_{\,\pm1.3}$ & 63.3$_{\,\pm4.5}$ & 81.7$_{\,\pm1.6}$ & 81.8$_{\,\pm3.2}$ \\
TKEO\,{+}\,HT & 63.2$_{\,\pm2.9}$ & 68.4$_{\,\pm3.3}$ & 81.0$_{\,\pm2.4}$ & 85.2$_{\,\pm2.7}$ \\
TKEO\,{+}\,DE & 62.6$_{\,\pm3.1}$ & 68.5$_{\,\pm3.3}$ & 80.5$_{\,\pm2.8}$ & 85.1$_{\,\pm2.8}$ \\
\hline
\end{tabular}
\label{tab:results_seed}
\end{center}
\vspace{-0.7cm}
\end{table}


\begin{table}[!h]
\fontsize{7.8pt}{8.8pt}\selectfont
\setlength{\tabcolsep}{3pt}
\renewcommand{\arraystretch}{0.95}
\captionsetup{skip=-2pt}
\caption{5-fold Mean Classification Performance and Standard Deviation on TUEP (Epilepsy Detection).}
\begin{center}
\begin{tabular}{|c|c|c|c|c|}
\hline
\multirow{2}{*}{\textbf{Features}} 
& \multicolumn{2}{|c|}{\textbf{Balanced Accuracy (\%)}} 
& \multicolumn{2}{|c|}{\textbf{ROC-AUC ($\mathbf{\times 10^2}$)}} \\
\cline{2-5}
 & \textbf{Raw} & \textbf{Fused} & \textbf{Raw} & \textbf{Fused} \\
\hline 
SE & 56.8$_{\,\pm2.5}$ & 62.9$_{\,\pm6.3}$ & 61.4$_{\,\pm11.0}$ & 69.3$_{\,\pm19.1}$ \\
PSD & \textbf{62.8$_{\,\pm5.8}$} & 62.4$_{\,\pm8.3}$ & \textbf{71.2$_{\,\pm10.9}$} & 71.2$_{\,\pm13.1}$ \\
HT & 57.2$_{\,\pm3.2}$ & 60.3$_{\,\pm4.3}$ & 67.3$_{\,\pm6.8}$ & 70.9$_{\,\pm15.5}$ \\
DE & 51.3$_{\,\pm2.0}$ & 53.2$_{\,\pm0.8}$ & 56.1$_{\,\pm4.5}$ & 64.8$_{\,\pm5.1}$ \\
\hline 
TKEO & 58.0$_{\,\pm7.3}$ & 62.4$_{\,\pm7.1}$ & 63.3$_{\,\pm16.4}$ & 69.6$_{\,\pm18.8}$  \\
TKEO\,{+}\,SE & 59.5$_{\,\pm5.2}$ & 61.7$_{\,\pm6.6}$ & 62.9$_{\,\pm12.9}$ & 67.1$_{\,\pm18.1}$ \\ 
TKEO\,{+}\,PSD & 62.3$_{\,\pm10.7}$ & \textbf{75.9$_{\,\pm8.0}$} & 70.2$_{\,\pm15.2}$ & \textbf{80.7$_{\,\pm11.4}$} \\
TKEO\,{+}\,HT & 62.7$_{\,\pm8.2}$ & 64.2$_{\,\pm6.9}$ & 69.0$_{\,\pm15.3}$ & 75.1$_{\,\pm18.3}$ \\
TKEO\,{+}\,DE & 59.7$_{\,\pm8.1}$ & 62.5$_{\,\pm6.9}$ & 64.5$_{\,\pm16.0}$ & 69.9$_{\,\pm18.2}$ \\
\hline
\end{tabular}
\label{tab:results_tuep}
\end{center}
\vspace{-0.3cm}
\end{table}

\begin{figure}[!h]
    \centering
    \includegraphics[width=0.88\linewidth]{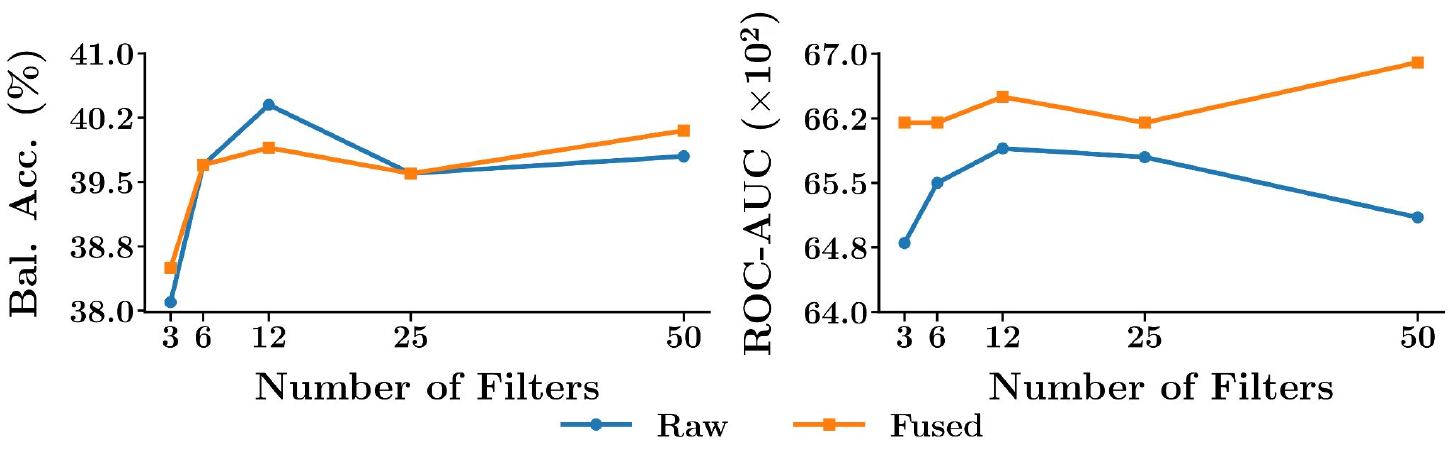}
    \caption{Performance of different filterbanks on BCI-IV\,2a.}
    \label{fig:ablation_filterbanks}
    \vspace{-0.5cm}
\end{figure}

\begin{figure}[!h]
    \centering
    \includegraphics[width=0.85\linewidth]{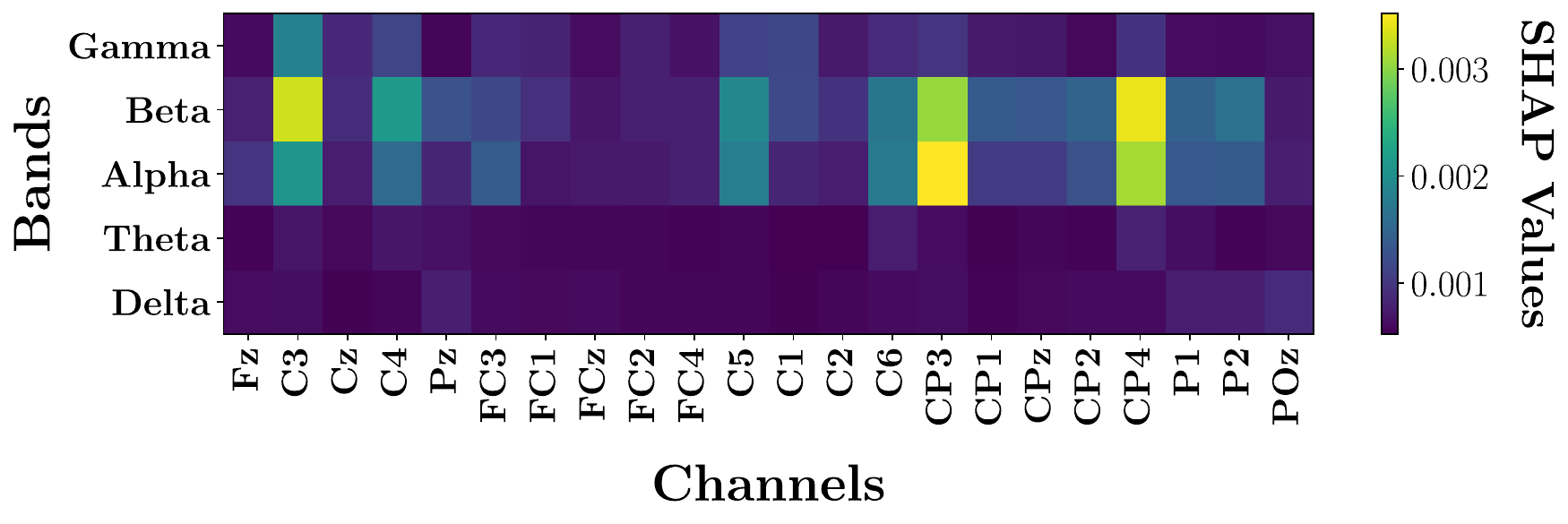}
    \caption{Absolute SHAP values of frequency bands for fused-band TKEO prediction contributions across EEG channels in BCI-IV\,2a; higher values indicate greater contribution.}
    \label{fig:shap}
    \vspace{-0.2cm}
\end{figure}

\textbf{Additional Experiments: }In our study, EEG signals were decomposed into frequency bands using a filterbank with a fixed number of filters. To examine the impact of filter count, we performed an ablation study on the BCI-IV\,2a dataset. As shown in Fig.~\ref{fig:ablation_filterbanks}, performance exhibited minor variations across configurations, with 12 filters achieving the highest accuracy for the raw-band and 50 for the fused-band. To further expand our study, we investigated the impact of each frequency band to the fused-band overall accuracies. As shown in Fig.~\ref{fig:shap}, the SHAP~\cite{shap} heatmap highlights elevated contributions in sensorimotor regions (C3,\,C4,\,CP3,\,CP4) in Alpha-Beta bands that capture event-related desynchronization/synchronization (ERD/ERS) during imagined movements. This pattern aligns with well-established neurophysiological findings of motor imagery~\cite{scherer2018}. We also evaluated the main pipeline restricted to sensorimotor\,regions\,(C3,\,Cz,\,C4,\,CP3,\,CPz,\,CP4) across different bands. The\,Alpha-Beta combination achieved the highest performance for this subset (40.9\% Balanced Accuracy, 67.4\% ROC-AUC; Table~\ref{tab:selected_channels}), indicating that these channels and bands retain discriminative information while reducing feature dimensionality.


\vspace{-0.2cm}

\begin{table}[!h]
\fontsize{7.6pt}{8.6pt}\selectfont
\setlength{\tabcolsep}{3pt}
\renewcommand{\arraystretch}{0.95}
\captionsetup{skip=-2pt}
\caption{Performance for Selected Channels on BCI-IV\,2a.}
\begin{center}
\begin{tabular}{|c|c|c|}
\hline
\textbf{Band} & \textbf{Balanced Accuracy (\%)} & \textbf{ROC-AUC ($\mathbf{\times 10^2}$)} \\
\hline
Raw & 39.3$_{\,\pm8.0}$ & 65.3$_{\,\pm7.8}$ \\
Alpha & 38.6$_{\,\pm9.3}$ & 64.7$_{\,\pm8.1}$ \\
Beta & 40.3$_{\,\pm8.2}$ & 66.8$_{\,\pm7.3}$ \\
Alpha\,{+}\,Beta & \textbf{40.9$_{\,\pm9.0}$} & \textbf{67.4$_{\,\pm8.2}$} \\
Fused & 40.6$_{\,\pm7.6}$ & 67.0$_{\,\pm7.3}$ \\
\hline
\end{tabular}
\label{tab:selected_channels}
\end{center}
\vspace{-0.8cm}
\end{table}


\section{Conclusion}
\label{sec:conclusion}
\vspace{-0.1cm}
In this paper, we investigated the utility of the Teager-Kaiser Energy Operator on three EEG classification tasks: motor imagery, emotion recognition, and epilepsy detection. A Gabor filterbank was applied within each canonical band to isolate subbands, followed by the application of TKEO to extract statistical features, and instantaneous modulation features using the DESA-1 demodulation algorithm. Experimental results demonstrate that TKEO-based features achieve competitive performance compared to baseline measures and contribute complementary information, improving accuracy in motor imagery and epilepsy detection tasks, while performing on par with energy-based features on emotion recognition. Feature fusion across frequency bands further improves performance, highlighting the narrowband sensitivity of TKEO. Overall, our analysis indicates that TKEO is a valuable tool for EEG signal analysis, both as a standalone method and as a complementary feature extraction technique. Future work could explore integrating TKEO into deep learning (DL) frameworks, combining the predictive power of DL with physiologically meaningful, interpretable representations. Another direction is combining TKEO-based features, which capture local transient energy dynamics, with functional connectivity measures for modeling large-scale neural interactions~\cite{fc}.


\vspace{-0.24cm}

\vspace{0.075cm}
\bibliographystyle{IEEEtran}
\bibliography{refs}

\end{document}